\documentclass[3p,review]{elsarticle}

\usepackage{lineno,hyperref}
\modulolinenumbers[5]

\usepackage{amssymb}
\usepackage{graphicx}
\usepackage{caption}
\usepackage{subcaption}
\usepackage{longtable}
\usepackage{multirow}
\usepackage[T1]{fontenc}
\usepackage{array}
\usepackage{booktabs}
\usepackage{rotating}
\usepackage{tabularx}

\usepackage[T1]{fontenc}

\usepackage{mwe}
\usepackage{chngpage}
\usepackage{lipsum}
\usepackage{booktabs}
\usepackage{xcolor}
\usepackage{enumitem,kantlipsum}
\journal{Neurocomputing}

\begin{document}

\begin{frontmatter}

\title{Medical Image Segmentation with 3D Convolutional Neural Networks: A Survey}

\author[A]{S Niyas  \corref{c-c21ebc19ae78}}
	\ead{niyasknpy@gmail.com }\cortext[c-c21ebc19ae78]{Corresponding author.}
	\author[A]{S J Pawan}
	\ead{pawansnj03@gmail.com }
	\author[B]{M Anand Kumar}
	\ead{m_anandkumar@nitk.edu.in}
    \author[A]{Jeny Rajan}
	\ead{jenyrajan@nitk.edu.in}
	
	\address[A]{Department of Computer Science and Engineering\unskip, 
		National Institute of Technology Karnataka\unskip, Surathkal\unskip, Mangalore – 575025\unskip, Karnataka\unskip, India.}
	\address[B]{Department of Information Technology\unskip, 
		National Institute of Technology Karnataka\unskip, Surathkal\unskip, Mangalore – 575025\unskip, Karnataka\unskip, India.}

\begin{abstract}
Computer-aided medical image analysis plays a significant role in assisting medical practitioners for expert clinical diagnosis and deciding the optimal treatment plan. At present, convolutional neural networks (CNNs) are the preferred choice for medical image analysis. In addition, with the rapid advancements in three-dimensional (3D) imaging systems and the availability of excellent hardware and software support to process large volumes of data, 3D deep learning methods are gaining popularity in medical image analysis. Here, we present an extensive review of the recently proposed 3D deep learning methods for medical image segmentation. Furthermore, the research gaps and future directions in 3D medical image segmentation are discussed.
\end{abstract}


\end{frontmatter}

\section{Introduction}
\label{intro}

Artificial intelligence (AI) techniques have been extensively used for analyzing medical images in the modern healthcare system. Computer-aided diagnosis (CADx) enables analyzing the medical data and extracting valuable information using specialized software to assist physicists in making rapid and informative clinical decisions. These diagnostic systems reduce the subjectivity in decision-making and the overall cost involved. In recent years, popular medical imaging modalities, such as X-ray, computed tomography (CT), ultrasonography (USG), and magnetic resonance imaging (MRI), have advanced in terms of acquisition time, image quality, and cost-effectiveness. \cite{chakraborty2018intelligent}. However, these techniques exhibit inherent limitations such as noise, motion artifacts, non-uniform contrasts, and registration errors, leading to an unreliable clinical diagnosis. Medical image processing entails problem-specific strategies based on image processing algorithms. Some of its typical applications include image registration, denoising, enhancement, compression, classification, and segmentation, which predominantly used traditional machine learning models to analyze the characteristics such as contrast variation, orientation, shape, and texture patterns. However, machine learning models exhibit various limitations \cite{Goodfellow-et-al-2016,shen2017deep}, such as 1) high dependency on handcrafted features extracted by a domain expert, 2) inevitable manual intervention, and 3) arbitrary parameters. These limitations of traditional machine learning algorithms have resulted in automated feature extraction methods such as convolutional neural networks (CNNs) \cite{abiodun2019comprehensive}.

\par
Studies on neural-network-based decision-making are being conducted since the 1950s. Several challenges were faced initially, mainly due to the lack of ample data and adequate computational capabilities. However, factors such as digitalization that lead to data availability, advancements in parallel distributed computing, and the surge in the semiconductor industry accelerated advanced research on neural networks. With the increasing prominence of image data in the digital data space, research on developing neural network models for image analysis increased tremendously, as evident from the introduction of convolutional neural networks (CNN). The first popular CNN method was published by LeCun et al. \cite{lecun1989backpropagation} for handwritten character recognition. Subsequently, several successful deep CNN models have been developed to solve various classification \cite{dhillon2020convolutional} and segmentation problems \cite{minaee2021image}.

\par
\begin{figure}[!t]
  \centering
  
  \begin{subfigure}{1.0\textwidth}
  \includegraphics[width=\textwidth]{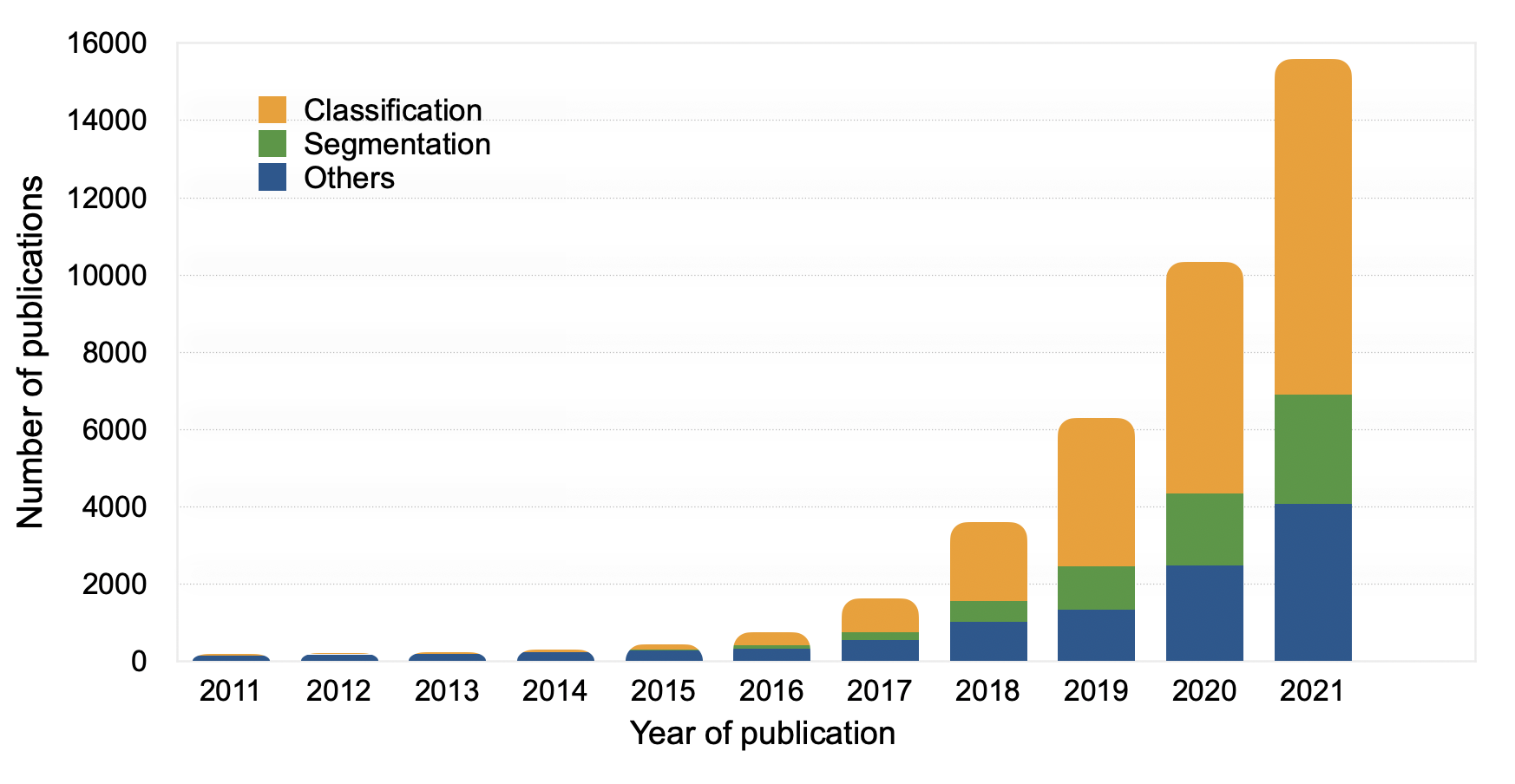}
  \caption{ Breakdown of papers based on the application.}
  \label{fig_pubmed1}
  \end{subfigure}

  \begin{subfigure}{0.97\textwidth}
  \includegraphics[width=\textwidth]{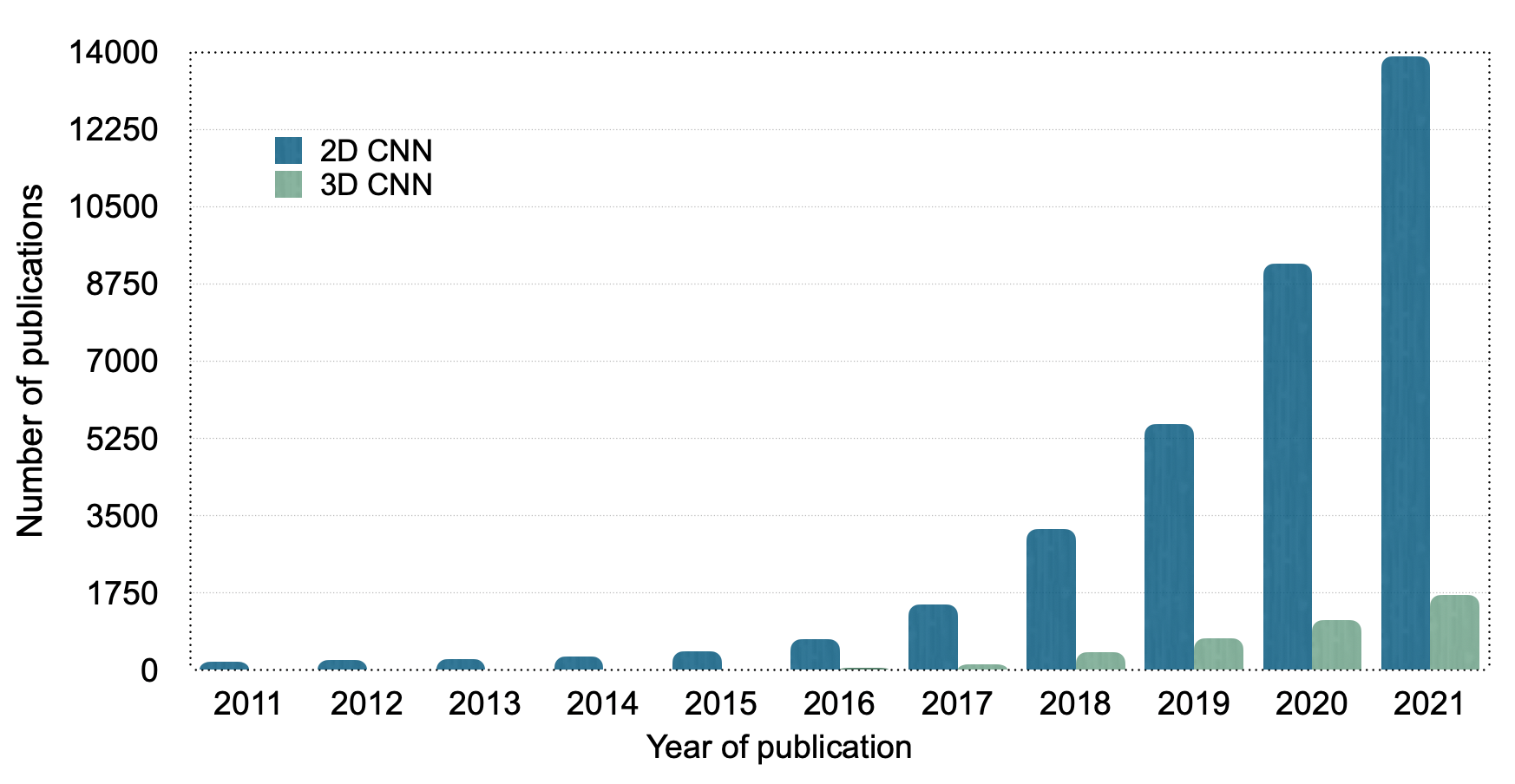}
  \caption{ Breakdown of papers based on the deep learning approach (2D or 3D CNN).}
  \label{fig_pubmed3}
  \end{subfigure}
  
  \caption{Statistics of papers retrieved from PubMed on medical image analysis using deep learning techniques (from 2011-2021). The data for the year 2021 has been extrapolated from the papers listed till June 2021.}
  \label{fig_pubmed}
\end{figure}

Medical image processing has also benefited from these developments periodically. As a result, several studies related to medical image analysis using deep learning techniques have been reported in the literature. Despite the significant growth in the research on deep learning-based medical image analysis during the last couple of years, statistics on classification outweigh that on segmentation approaches. The statistics of CNN-based medical image processing papers retrieved from PubMed are shown in Fig.~\ref{fig_pubmed1}.

\par
The literature reports numerous dedicated reviews on CNN-based medical image analysis. For instance, the review article by Shen et al. \cite{shen2017deep} provides a comprehensive overview of medical image processing using advanced machine learning and deep-learning techniques. Their study highlights the advantages of various CNN models over traditional machine learning methods in different medical applications. Similarly, Litjens et al. \cite{litjens2017survey} thoroughly discussed the applications of deep learning in various fields of medical image analysis. In addition, several review papers \cite{yasaka2018deep,sahiner2019deep,currie2019machine,ravi2016deep,yamashita2018convolutional,ker2017deep} provide a basic understanding of CNN models and their applications in medical image analysis. However, most of these reviews have focused on analyzing deep learning approaches in an application-specific manner.

\par
Since CNN models are designed to operate using image data, the dimensionality of the images plays a significant role in selecting the appropriate model. Generally, the majority of the medical image data are constructed using either two-dimensional (2D) or 3D imaging techniques. For example, digital X-rays, retinal fundus images, microscopic images in pathology, mammograms, etc., belong to the 2D image categories in the biomedical domain. Similarly, MRI, CT, and ultrasound (US) are widely used 3D medical images. However, most of the published studies on 3D medical image analysis have used 2D CNN models. Fig.~\ref{fig_pubmed3} substantiates our statement showing the breakdown of studies that have used 2D and 3D deep learning techniques in medical image analysis. 

\par
Segmentation of organs or anatomical structures is a crucial step in medical image processing. It is primarily used to detect abnormalities and estimate the true extent of the organ or lesion. In \cite{singh20203d}, Singh et al. presented an overview of various building blocks in 3D CNN architectures and several deep-learning approaches in volumetric medical image analysis. To the best of our knowledge, there are no comprehensive reviews in the literature focusing on medical image segmentation using 3D deep learning techniques. This motivated us to conduct an in-depth review of current deep learning trends in 3D medical image segmentation. We have also discussed the future perspectives in 3D medical image segmentation.

\par
This review covers several articles that mainly focused on the 3D deep-learning techniques for volumetric medical image segmentation. Papers reporting overlapping techniques have been excluded from this review to minimize redundancy. The primary contributions of this review article are to provide
\begin{enumerate}
    \item an overview of 3D deep learning for medical image processing.
    \item an in-depth review on various state-of-the-art 3D deep learning-based medical image segmentation.
    \item a discussion on research gaps and future directions in 3D medical image segmentation.
\end{enumerate}

\par
The rest of this review is organized as follows. Section \ref{section2} presents a brief overview of 2D and 3D CNN models and a detailed survey on existing 3D deep learning models along with their contributions to medical image segmentation. Section \ref{section4} provides a critical discussion and an outlook for future research.

\section{3D CNN in Medical Image Segmentation}
\label{section2}

\subsection{Convolutional Neural Networks: An overview}

\par
Convolutional neural networks are the most popular machine learning methods at present. In CNN models, a set of kernels (or filters) are utilized to learn various characteristics, and these learnable kernel values are updated based on the value of the cost function. The convoluted output from each kernel is considered as the feature map and is passed to the subsequent layers. While using a set of $K$ kernels, the feature map generated in the $n^{th}$ layer can be represented as:
\begin{equation}
{X_{k}}^{n} = \sigma ({w_{k}}^{n} * {X}^{n-1} + {b_{k}}^{n}), k = \{1,2,..., K\}
\label{eq:CNN1}
\end{equation}
where ${w_{k}}^{n}$ is the $k^{th}$ kernel in the $n^{th}$ layer, and ${b_{k}}^{n}$ is the $k^{th}$ bias in the $n^{th}$ layer. ${X_{k}}^{n}$ represents the feature map created in the $n^{th}$ layer, $X^{n-1}$ is the feature map from ${(n-1)}^{th}$ layer, and $\sigma (\cdot)$ represents the activation function.

\par
CNNs do not require separate filters to detect similar objects from multiple locations in the image and hence make the number of weight parameters independent of the image size. This helps to reduce the number of trainable parameters and makes the operation translational invariant across the image. CNN architecture (for classification) generally consists of three building blocks: convolution layers, pooling layers, and fully connected dense layers \cite{yamashita2018convolutional}. A convolution layer uses small multi-dimensional kernels to extract spatial features from image locations. The pooling layer is designed to pass relevant features to the subsequent layers by down-sampling the feature space. In addition, pooling reduces the number of trainable parameters in the subsequent dense layers, expands the receptive field for multi-scale feature extraction, and provides translation invariance to small shifts.

\par
\begin{figure} [!htb]
\centering
\includegraphics[width=0.9\textwidth]{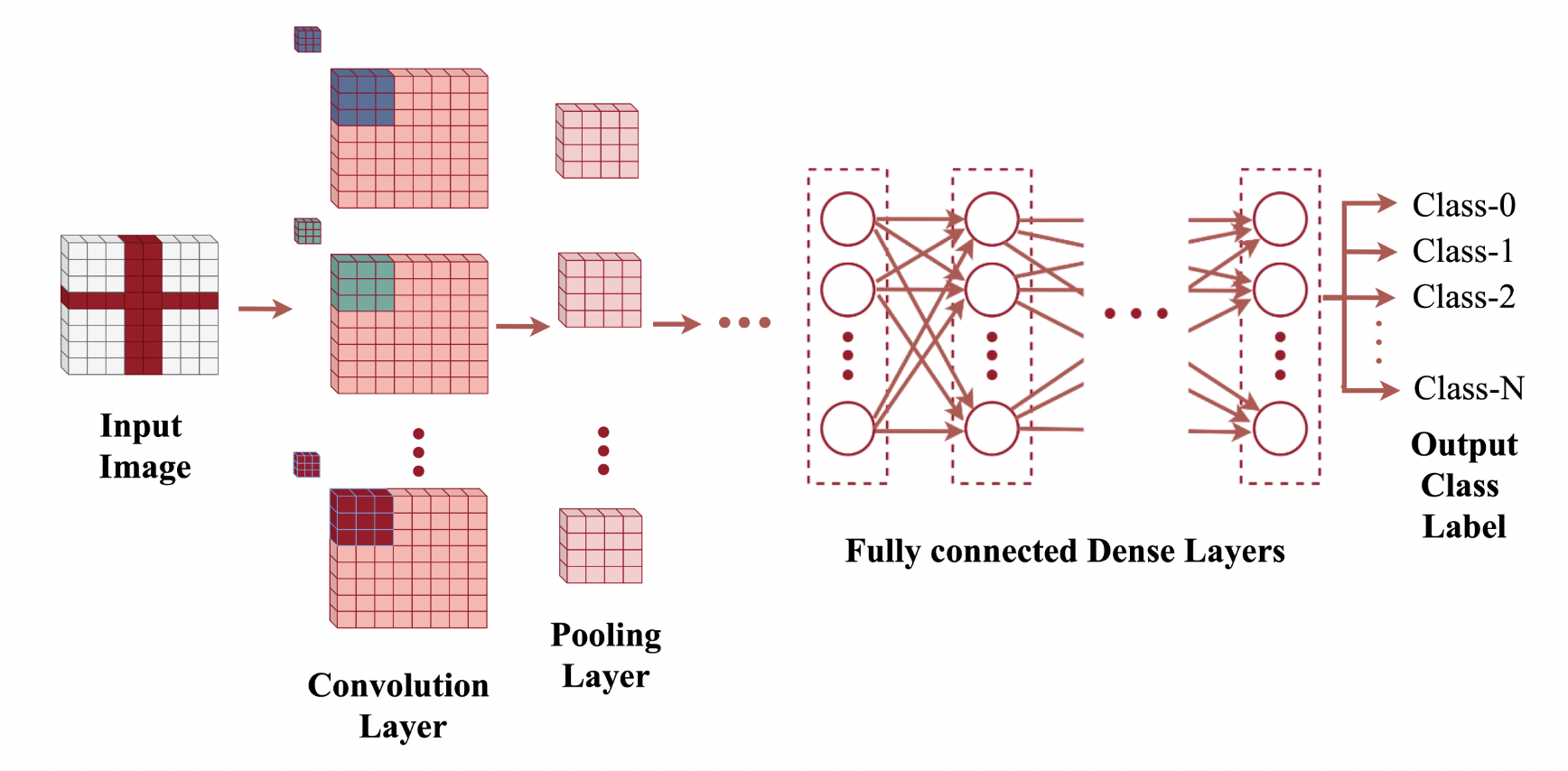}
\caption{A high level representation of a CNN classification model.}
\label{fig:CNN2}   
\end{figure}

A typical CNN model uses multiple convolution operations followed by activation and pooling layers for extracting multi-scale features; the complexity of image features advances as the model depth increases. The fully connected dense layer is designed to transform the feature map from the previous layer to one-dimensional feature vectors. The final fully connected layer operates as a classification layer and provides the probabilities of the target classification task. Furthermore, a CNN model uses several additional modules such as batch normalization \cite{ioffe2015batch}, regularization, and dropout \cite{srivastava2014dropout}, to improve the learning process \cite{singh20203d}. A high level representation of a CNN classification model is shown in Figure \ref{fig:CNN2}.

\subsection{2D CNN vs 3D CNN in Image Segmentation}

In image segmentation, every pixels needs to be classified simultaneously. CNN model for segmentation uses convolution and pooling layers similar to those in a classification network. However, the fully connected layers are usually excluded in the segmentation model to retain the spatial relationship of pixels throughout the network. In addition, adequate upsampling layers are added to the final layers of the network to compensate for the pooling operation. Hence, the probability of each pixel is estimated and finally, the segmentation mask is created.

\par
Long et al. \cite{long2015fully} proposed the first fully convolutional neural network (FCNN), which consists of a sequence of convolution and pooling layers to extract features from different scales of image. To compensate the reduction in feature space due to pooling, the latent space is upsampled directly to the actual image dimension using deconvolution and skip connections. Followed by this FCNN approach, several segmentation methods have been proposed. Liu et al. \cite{liu2015parsenet} improved the FCNN model by introducing global pooling, and Noh et al. \cite{noh2015learning} improved it by introducing a symmetric encoder-decoder network that uses VGG-16 Net layers in the encoder side. However, these early deep learning based segmentation approaches lack precision in the segmentation outcome and often miss fine details while reconstructing the segmentation mask from the latent feature space.

\par
\begin{figure}
    \centering
     \begin{subfigure}[b]{1.0\textwidth}
         \centering
         \centerline{\includegraphics[width=0.9\textwidth]{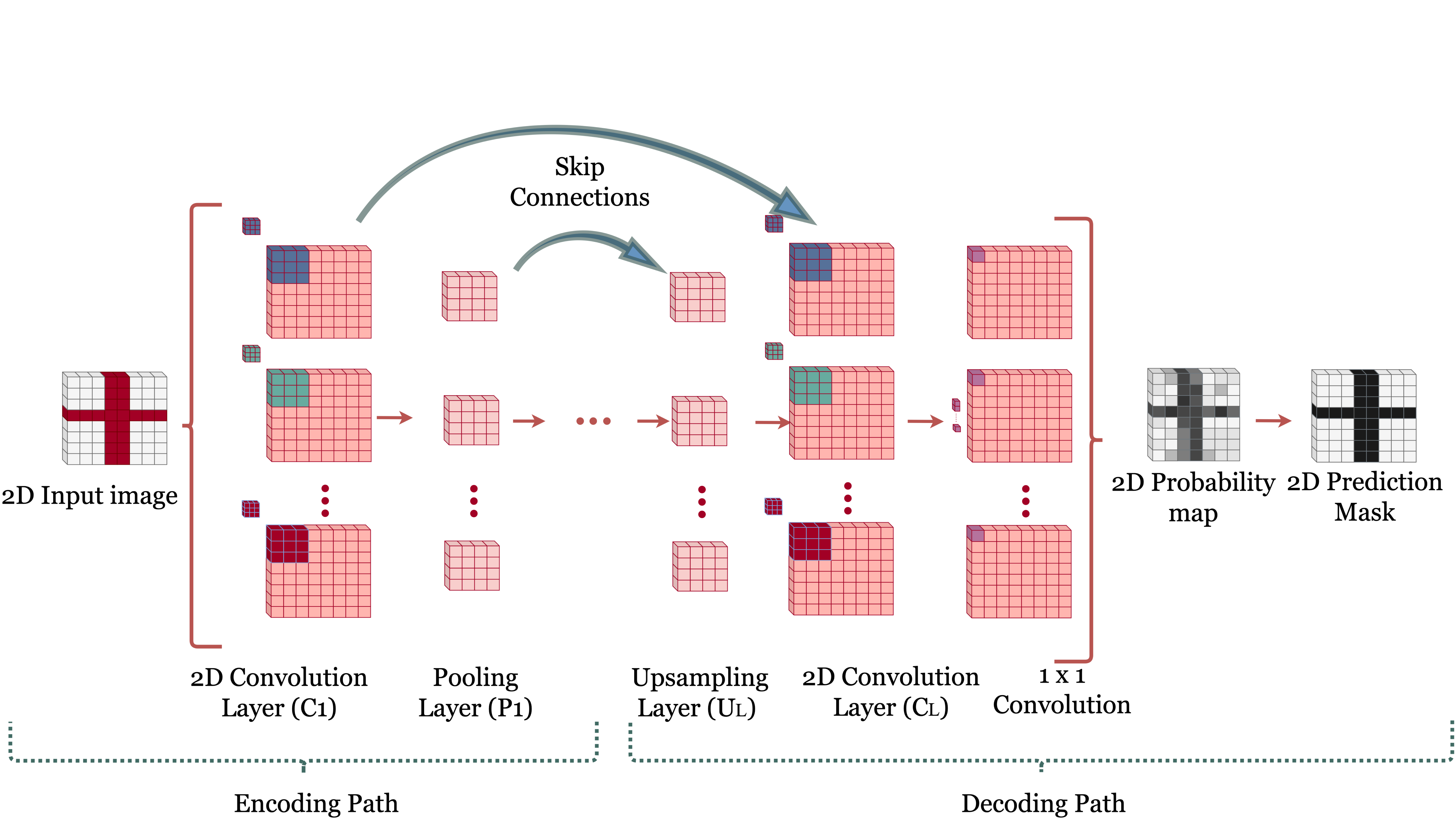}}
         \subcaption{A 2D CNN architecture for segmentation.}
         \label{CNN_seg}
     \end{subfigure}
     \hfill
     
     \centering
     \begin{subfigure}[b]{1.0\textwidth}
         \centering
         \centerline{\includegraphics[width=0.9\textwidth]{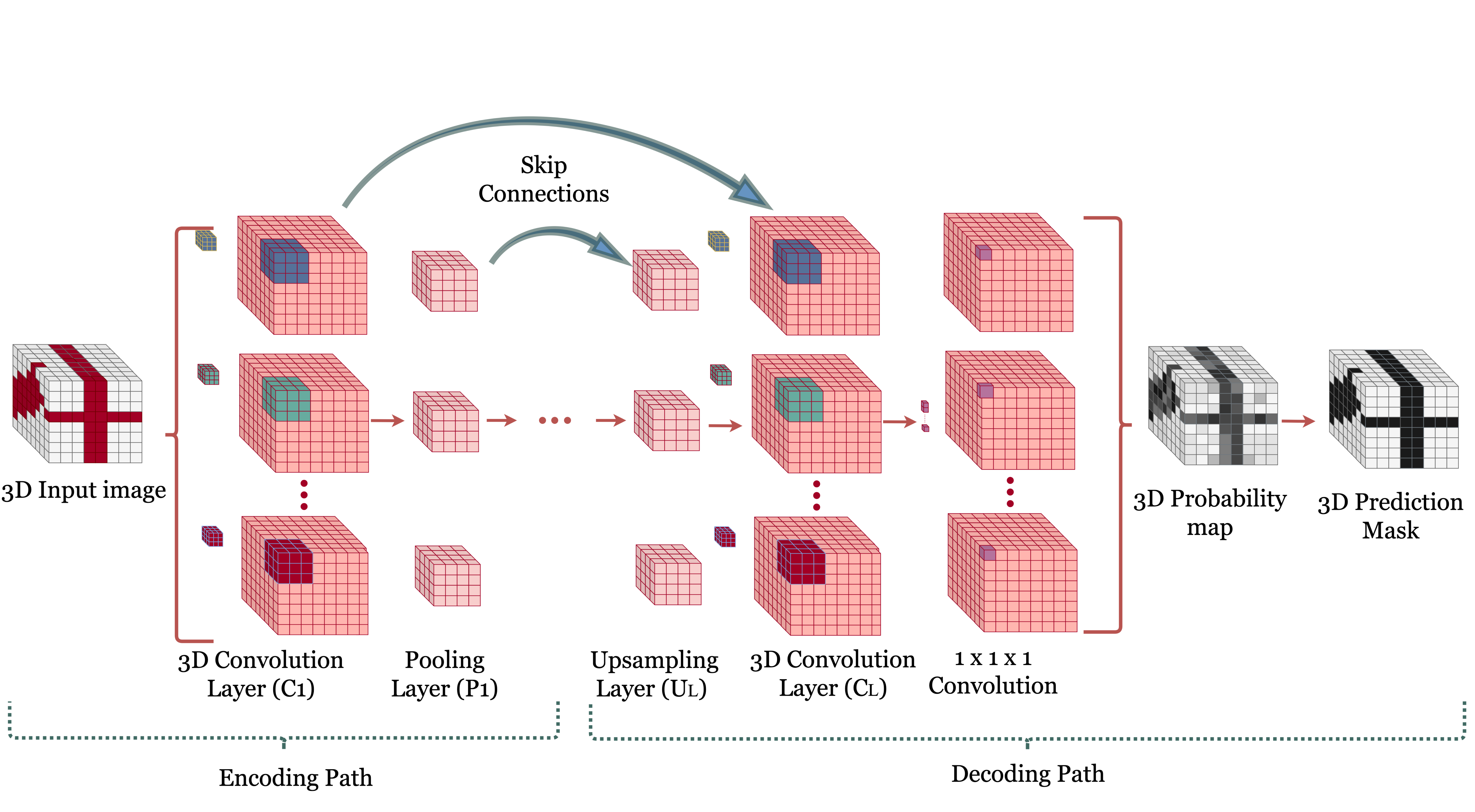}}
         \subcaption{A 3D CNN architecture for segmentation.} 
         \label{3DCNN_seg}
     \end{subfigure}
     \hfill
     
     \caption{2D and 3D CNN architectures for segmentation.}
\end{figure}

Ronneberger et al. \cite{ronneberger2015u} proposed an idea that can efficiently reconstruct the low-resolution latent vector to full-size images with the help of already learned features from the encoder layers. The network uses deconvolution in symmetry with the convolution layers to form an encoder-decoder architecture with skip connections to reduce the information loss and improve the saliency in semantic segmentation. Several improved versions of the semantic segmentation came later by designing the architecture with advanced convolution modules and design improvements. Recurrent neural networks (RNN) \cite{chen2016combining,chakravarty2018race}, DeepLab networks \cite{chen2014semantic,chen2017deeplab,chen2018encoder,chen2017rethinking}, Attention-based models \cite{nie2018asdnet,oktay2018attention}, and Generative Adversarial Network (GAN) based models \cite{yi2019generative,aggarwal2021generative} are some of the popular semantic segmentation methods. Although several CNN-based segmentation architectures have been proposed in the literature, the encoder-decoder based FCNN architectures and their derivatives \cite{ronneberger2015u,chen2017deeplab,chen2018encoder} are considered the best performing models for several segmentation problems. A general workflow of an encoding–decoding architecture for segmentation is depicted in Figure \ref{CNN_seg}. 

\par
A 3D CNN-based segmentation model uses 3D images as input, and a similar-sized 3D prediction mask is expected as the output. The network structure is similar to that of a standard 2D CNN model, but the convolution and pooling layers use 3D extensions to process the volumetric data. Here, the convolution layers perform filtering with 3D kernels, and the 3D pooling layers subsample the data in all three dimensions to compress the size of the feature space. Hence the volumetric data is analyzed as cubic patches from layer to layer and can learn spatial features in all three dimensions.

\par
Various 3D CNN based segmentation models have been proposed in the literature. For the sake of simplicity, the following explanation considers the popular encoder-decoder architecture as shown in Figure \ref{3DCNN_seg} In the encoding path, 3D convolution layers and pooling layers create 4D feature space. The number of filters used in each convolution layer determines the extent of the fourth dimension. Hence, the intermediate convolution layers use 4D feature maps. Since the pooling reduces the feature space dimension in the encoding path, the decoding path uses a sufficient number of upsampling layers to gradually increase the feature space dimension similar to the input data. Normally, transpose convolution is used for upsampling the data in a learnable fashion. Skip connections are also used to preserve the fine features, by merging features between the corresponding encoding and decoding layers.  A $(1\times1\times1)$ filter is used in the final layer to project the stacked features into a feature space with the same dimension as the input image. The probability map is then created using a non-linear activation function (commonly using the Softmax \cite{nwankpa2018activation} activation function to limit the probability values between 0 and 1), followed by thresholding which creates the final prediction mask. The key differences between the 2D and 3D deep learning methods for image segmentation are listed below.

\begin{enumerate}
    \item In 2D CNN, the convolution kernel moves in two (x and y) directions, while in 3D convolution, it moves in three (x, y, and z) directions. 
    \item 2D convolution can learn only 2D spatial features while 3D convolution can also extract inter-slice information from adjacent frames.
    \item In 2D CNN, the model accepts 2D images (or matrices) as the input data and provides corresponding 2D segmentation maps. In 3D CNN, both the input and output are 3D in nature.
    \item The input and output feature space of 2D convolution layers (except in the input and final layers) are 3D in nature. Similarly, the input and output feature space of 3D convolution layers are 4D.
    \item Since the convolution operations in a 3D CNN are significantly higher, and it needs more memory space to save the parameters and feature space than 2D CNN. 
\end{enumerate}

\subsection{An Overview on 3D Medical Data}

Medical images are acquired with different imaging principles. The characteristics of these images differ, in terms of spatial resolution, image intensity range, size of the image, and noise. In addition, they vary in terms of dimensionality and data representation. 2D data mostly use relatively simple Euclidean representation, which describes each data point using pixel intensity values. In medical images, these pixel values represent the state of the anatomical structure under consideration. For instance, the pixel intensity of X-ray images varies with the radiation absorption, whereas it depends on the acoustic pressure in ultrasound images or the radio frequency (RF) signal amplitude in MRI.

\par
The representation of 3D data is more complex. Nevertheless, several standard representations are reported in the literature, such as projections, volumetric representations, point clouds, and graphs \cite{ahmed2018survey}. The majority of the 3D medical image data belong to the volumetric representation, where volumetric elements or \emph{voxels} (analogous to \emph{pixels} in a 2D image) are used to model 3D data by describing how the 3D object is distributed through the three perpendicular axes. While acquiring 3D medical data, the device scans the body parts in any of the three planes: axial (front to back), coronal (top to bottom), and sagittal (side to side). Normally, multiple 2D slices are acquired across the area under consideration and are stacked to produce 3D image volumes using appropriate image registration techniques \cite{cox1999real,hajnal2001medical,sakas2002trends}.

\par
The 3D image visualization has provided a great opportunity for clinicians to evaluate the cross-section of anatomic structures. This has increased the understanding of the complex patterns and structural morphology, mostly in radiology. Currently, 3D imaging is commonly used in several modalities such as CT, MRI, USG, and PET. Although 3D imaging techniques have numerous advantages over 2D images, they have certain limitations as well. For example, compared to 2D imaging methods, they require significantly large storage space and are often expensive. However, over time visualization and interpretation of 3D medical data have become simple, thanks to modern deep learning algorithms and the availability of powerful graphics processing units (GPUs) \cite{zhou2020review}.

\subsection{Why 3D CNN is Important in 3D Medical Image Segmentation?}

Several medical image segmentation approaches proposed in the literature use 2D deep learning methods \cite{liu2021review,roth2018deep}. Though 2D CNN can learn the spatial relationship between the pixels in a 2D plane, it cannot learn the inter-slice relationship between the frames. In medical images such as X-ray or Fundus images, information is restricted into a 2D plane, and hence 2D convolutions can effectively detect relevant features. However, in 3D volumetric data such as MRI, CT, or USG, the region of interest may spread across multiple frames, and hence the inter-slice information became significant. In such cases, individual slice analysis with 2D convolutions cannot retrieve all useful information and may affect the segmentation performance. On the other side, 3D CNN uses convolution kernels in three directions, and hence the inter-slice information can be learned to provide better segmentation results. For instance, Table \ref{computation2} shows a comparison of 2D and 3D CNN in terms of computation complexity and memory requirements while segmenting brain tumor regions from MR images.\cite{menze2014multimodal}.

\par
The 2D and 3D UNet models used in this analysis are identical except in the convolution and pooling operation. The study uses a 3D dataset containing 224 samples, each sample having a size of $256\times256\times32$. The same dataset is used in 2D CNN after reshaping, consisting of 7168 2D samples, each having a size of $256\times256$. This ensures that the total number of frames used for training is identical for both 2D CNN and 3D CNN models. The number of filters in each convolution layer is designed to make the trainable parameters comparable for both models. The training time required to complete one epoch in the 3D CNN model is almost half the time required in the 2D CNN model. This advantage in computation time in 3D CNN is due to the optimization in vector multiplication while using more parameters in a single 3D convolution kernel. However, in practice, 3D CNN may require a large number of samples to achieve high segmentation accuracy, which increases the computation complexity. Since 3D CNNs have to be processed on 4D feature maps, the graphical memory requirement is also higher than a 2D network, and can be preferred over 2D CNN models when the inter-slice information is crucial and hardware resources with high performance computing are available.

\par
\begin{table}[]
\caption{Performance comparison of 2D and 3D UNet, in terms of number of trainable parameters ($T_{N}$), training time per epoch $T_{T}$ and GPU memory required $T_{M}$.}
\centering
\begin{tabular}{@{}cccccc@{}}
\toprule
CNN Models & \begin{tabular}[c]{@{}c@{}}No. of samples\\  used for training \end{tabular} & {\begin{tabular}[c]{@{}c@{}}Sample \\ size\end{tabular}} & {\begin{tabular}[c]{@{}c@{}}$T_{N}$\\ (in million)\end {tabular}} & {\begin{tabular}[c]{@{}c@{}}$T_{T}$ \\ (in seconds)\end{tabular}} & {\begin{tabular}[c]{@{}c@{}}$T_{M}$\\ (in GB)\end{tabular}} \\ \midrule
{3D UNet} & 224 & $256\times256\times32$ & 1.656 & 52 & 2.22 \\
{2D UNet} & 7168 & $256\times256$ & 1.692 & 108 & 0.176 \\ \bottomrule
\end{tabular}
\label{computation2}
\end{table}

\subsection{3D CNN in Medical Image Segmentation}

3D CNNs have been successfully used in several medical image segmentation tasks and classified into various subgroups based on different learning approaches. In this study, we considered 3D CNN papers for medical image segmentation published between 2015 and 2021 and are classified into one of the following categories:

\begin{enumerate}
    \item Fully supervised 3D CNN models
    \begin{enumerate}
        \item Direct 3D CNN models
        \item 3D Patch-wise segmentation models
        \item Multi-task learning models
    \end{enumerate}
    \item 3D CNN with Semi-supervised learning
    \item 3D CNN with Weakly-supervised learning
    \item Cost-effective approximations of 3D CNN
\end{enumerate}

Here the main categorization is based on whether or not the model is learning from fully annotated 3D data. The fully supervised 3D CNN uses image volumes with corresponding labeled masks for the training process. Many works in the literature belong to this category as the approach is straightforward with less complexity in the training process \cite{litjens2017survey,liu2021review}. Based on the implementation of the 3D CNN architecture and the nature of the learning paradigm, they are further classified into three main sub-classes. More explanation of these models are given in section \ref{1}. Since the labeling of large medical image datasets is challenging, research also expanded to deep learning models that can learn from limited labeled samples. In recent years, a few 3D CNN papers were also proposed in these semi-supervised and weakly-supervised medical image segmentation tasks and are included in the current literature review.

\par
3D CNN models demand substantially more memory, and various studies have been published that can simulate 3D CNN in order to extract inter-slice features using cost-effective approximations \cite{Ciompi2017TowardsAP,Mlynarski20193DCN,rickmann2020recalibrating}. This survey also discusses a few of these efforts in order to get a better understanding of 3D medical image segmentation. The subsequent sections performs a detailed review and provides in-depth comparison of state-of-the-art 3D medical image segmentation methods within each category.

\subsubsection{Fully supervised 3D CNN models}
\label{1}
\par

\begin{enumerate}
    \item \textit{Direct 3D CNN models }
    \par
A straightforward implementation of a 3D CNN model is possible by replacing the 2D convolution and pooling operations in a conventional CNN model with 3D convolution and pooling. A basic representation of such a 3D architecture is shown in Figure \ref{3DCNN_seg}. Such CNN models that use 3D equivalents of existing 2D segmentation architectures are reviewed under this category.

\par
\hspace{0.5cm} Milletari et al. \cite{milletari2016v} presented an advanced  3D U-Net architecture (V-Net) with residual blocks (instead of cascaded convolution blocks) and strided convolution (instead of max-pooling). The methodology use a Dice score based loss function to reduce the class imbalance between the voxel classes. This model was evaluated on the PROMISE2012 dataset \cite{litjens2014evaluation}, and the results were very close to the state-of-the-art 2D segmentation models. Bui et al. \cite{bui20173d} proposed another 3D CNN architecture inspired by the 2D FCNN model proposed by Long et al. \cite{long2015fully} for brain segmentation in infant brain MRI volumes \cite{wang2019benchmark}. The encoder path uses multiple coarse and fine dense 3D convolution layers to extract multi-scale features. Downsampling makes use of strided convolution to reduce the feature size and to increase the receptive field. Multiple convolution layers are used to extract four different scales of feature space, then upsampled and concatenated to generate the final probability map. However, direct merging of the upsampled features in the decoder path creates semantic gaps in the feature space, and may create undesired outcomes in the segmentation results. 

\par
\hspace{0.5cm} In \cite{kayalibay2017cnn}, Kayalibay et al. proposed another 3D encoder-decoder architecture with deep supervision \cite{wang2015training}. In this model, the feature space created at different decoder levels in the network are merged using an element-wise summation of the features. Hence, the learning is based on the error between the ground truth and a combination of feature maps from different decoding levels. In the approach, the learning process is directly dependent on the coarse features from the different decoder levels, and helps to speed-up the convergence. However, the element-wise summation may limit the learning process when the semantic gap between the different decoder levels is significant. Dou et al. \cite{dou20173d} proposed a similar approach that can work on relatively small 3D datasets. The model uses an encoder-decoder FCNN with 3D convolutions. The outputs from each decoder layer are upsampled using 3D deconvolution and merged to get the final segmentation mask. The integration of deep supervision and the post-processing using CRF \cite{zheng2015conditional}  helped to improve the overall segmentation performance.

\par
\hspace{0.5cm}In \cite{li2017compactness}, Li et al. presented a 3D CNN model using dilated convolution \cite{wei2018revisiting} instead of max-pooling, and residual connections. For utilizing multi-scale features, the dilation factor of the 3D convolution kernel is steadily increased in the subsequent layers. Hence, the spatial resolution of the feature space is kept constant throughout the network. The residual blocks merge the features from different layers to reduce vanishing gradients problem and feature degradation. However, the absence of max-pooling reduces the translation invariance in the feature space and accelerates the computation complexity. 

\par
\hspace{0.5cm}Chen et al. \cite{chen2018voxresnet} proposed a residual deep 3D CNN architecture for segmenting the brain region from 3D MRI volumes. The proposed architecture employs a voxel-wise residual network (VoxResNet). The architecture uses a 3D extension of 2D deep residual networks that extracts features from multiple scales using a series of convolution operations, and the features at multiple scales are fused to generate the segmentation mask. For training such a deep network with small training data, multi-modal contextual information is integrated into the network to take advantage of additional information from various MRI sequences. Experiments were conducted using three MRI sequences: T1, T1-IR, and T2-FLAIR from the brain structure segmentation task \cite{mendrik2015mrbrains}.

\par
\hspace{0.5cm}In \cite{niyas_3DFCD}, Niyas et al. presented a 3D Residual U-Net architecture for segmenting focal cortical dysplasia (FCD) from 3D brain MRI. The method aims to retain the advantages of both 2D and 3D CNN methods by an effective design of input data slices and CNN architecture. The model proposes a shallow sliced stacking approach to generate large number of 3D samples from small datasets. The customized 3D U-Net architecture with residual connections in the encoder path helps in extracting multi-scale features with fewer trainable parameters.

\par
\hspace{0.5cm}Schlemper et al. \cite{Schlemper2018AttentionUL} proposed an attention gate (AG) based 3D U-Net architecture for medical image segmentation which automatically learns to concentrate on various target structures. The features from the encoding path to the decoding path are propagated through the skip connections with self-attention gating modules \cite{Hu_2018_CVPR}. The AG module uses grid-based gating that allows attention coefficients to focus more on local neighborhoods, that helps to locate the relevant regions in the image. Performance improvements of the proposed network over U-Net are experimentally observed to be consistent across different imaging datasets \cite{p26}, \cite{Roth2017Hierarchical3F}. However, in an FCNN with encoder-decoder architecture, features in the lower depth are minimal and naive, while the features at advanced depths are of higher granularity. Hence, the attention modules at shallow depths may not help passing the salient features to the upsampling layers \cite{Thomas2020MultiResAttentionU}. 

\par
\hspace{0.5cm}Wang et al. \cite{Wang2019RPNetA3} proposed another 3D FCNN method that integrates recursive residual blocks and pyramid pooling to extract more complex features. The recursive residual blocks contain multiple residual connections that can minimize feature degradation problems. Pyramid pooling \cite{He2015SpatialPP} generates fused feature maps at different decoding levels for obtaining both local and global information. It helps to eliminate the fixed size constraints of CNN without losing spatial information. The proposed architecture gave better multi-class segmentation performance while detecting white matter (WM), gray matter (GM), and cerebrospinal fluid (CSF) from 3D MR images from CANDI, IBSR18, and IBSR20 datasets \cite{Kennedy2011CANDIShareAR,Rohlfing2004EvaluationOA}.

\par
\hspace{0.5cm}A 3D version of multi-scale U-Net segmentation was presented in \cite{Peng2020MultiScale3U} by Peng et al. The model uses multiple U-Net modules to extract long-distance spatial information at different scales. The U-Net blocks use Xception \cite{Chollet2017XceptionDL} modules instead of normal convolution to extract more complex features. Feature maps at different resolutions are upsampled and fused to generate the segmentation mask. The authors claimed that the upsampled feature maps at different scales could extract and utilize appropriate features with faster learning.  Further, the 3D convolutions are replaced with depth-wise separable convolutions to reduce the computation complexity. However, the complex structure of the overall network and the deep convolution layers demand high memory and computation requirements. \par

\par
\hspace{0.5cm}The model cascade (MC) strategy is one of the popular strategy in CNN-based image segmentation that can alleviate the class imbalance problem by using a set of individual networks for coarse-to-fine segmentation. Despite its notable performance, it leads to undesired system complexity and overlook the correlation among the models. Zhou et al. \cite{Zhou2020OnePassMN} proposed a lightweight deep 3D CNN architecture: one-pass multi-task network (OM-Net) that can handle the native flaws in the MC approach. OM-Net combines discrete segmentation tasks into a one-pass deep model, to learn joint features and solve the class imbalance problem better than MC. The prediction results between tasks are correlated using a cross-task guided attention (CGA) block that can adaptively re-calibrate channel-wise feature responses. The model shows significant performance improvements in multiple public datasets \cite{menze2014multimodal,bakas2017segmentation}. 

\par
\hspace{0.5cm}In the above-mentioned medical image segmentation approaches, the 3D CNN models mainly use a straight-forward extension of various 2D CNN building blocks. However, the volumetric data analysis brings a significant increase in the memory requirement and computation costs while training deep 3D CNN models compared to corresponding 2D versions. Several studies were also conducted in developing 3D segmentation models that works with limited training data and hardware resources. Such approaches are discussed in the following sections.

\item \textit{3D Patch-wise segmentation models}

The resolution of input 3D data is the primary reason behind the need for large memory and higher computation complexity in training a 3D CNN segmentation model. However, this could be circumvented with a patch-based approach. In patch-wise approach, the input 3D images are converted to small 3D sub-samples and analyzed individually. Finally, the patch-wise prediction outcomes will be merged to create the final 3D prediction map. A high-level block diagram of 3D Patch-wise segmentation is shown in Figure \ref{fig:Patch}. The patch-wise analysis is an interesting are of research in 3D medical image analysis, and such patch-wise 3D deep learning based techniques are discussed in this section.

\begin{figure} [!tbh]
\centering
\includegraphics[width=0.85\textwidth]{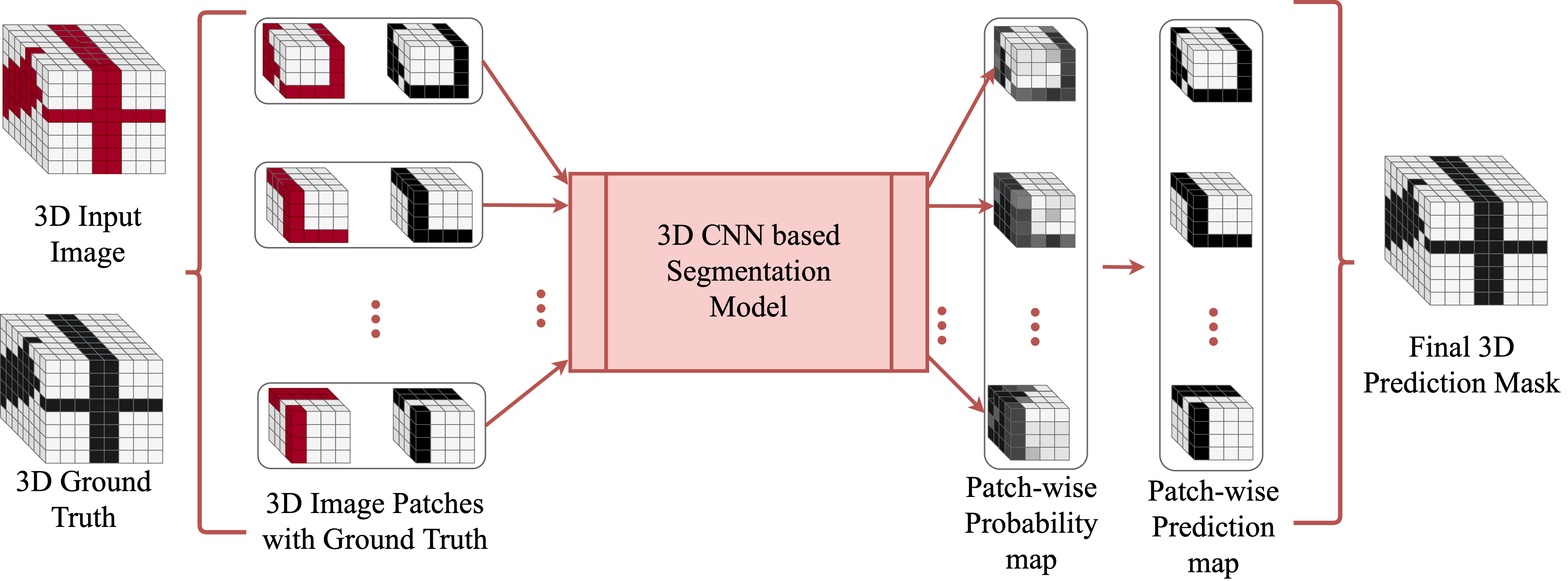}
\caption{A high-level block diagram of 3D Patch-wise segmentation model.}
\label{fig:Patch}   
\end{figure}

\par
\hspace{0.5cm}Yu et al.\cite{yu20163d} proposed a deep supervised 3D fractal network for whole heart and great vessel segmentation in MRI volumes. This method is an expansion of the FractalNet \cite{larsson2016fractalnet} and uses deep supervision using multiple auxiliary classifiers deployed at expanding layers to reduce the vanishing gradient problem. The methodology uses cropped 3D patches of size $(64\times64\times64)$ and reported good results in the HVSMR 2016 challenge dataset \cite{pace2015interactive}.
 
\par
\hspace{0.5cm}A deep dual pathway 3D CNN architecture was proposed by Kamnitsas et al. \cite{kamnitsas2017efficient} for brain lesion segmentation. They employed a two-way network that simultaneously learns from multiple image scales to analyze features from different receptive fields. The approach uses 3D patches at two different scales and classifies the center voxels as any target classes using fully connected dense layers. A 3D fully connected conditional random field (CRF) is also utilized in the post-processing stage to reduce false alarms in the final segmentation mask. This method shows improved segmentation results over BRATS 2015 \cite{menze2014multimodal} and ISLES \cite{halme2015isles} datasets, compared with various 2D CNN methods. An advanced version of this model was also proposed in \cite{kamnitsas2016deepmedic}, which used multiple residual connections in the encoding path. The model reported better segmentation performance and was bench-marked at BRATS 2016 challenge \cite{bakas2015glistrboost}.  

\par
\hspace{0.5cm}In \cite{kamnitsas2017ensembles}, Kamnitsas et al. presented an ensembles of multiple models and architectures (EMMA) for robust performance by aggregating predictions from multiple models. This method use an ensemble of a DeepMedic model \cite{kamnitsas2016deepmedic}, three 3D FCNN models \cite{long2015fully}, and two 3D U-Net models \cite{ronneberger2015u}. An ensembler computes the confidence maps, and finds the average class confidence of the individual models. Finally, each voxel in the segmentation map was labeled with the class label with the highest confidence score. The EMMA approach won the first place in the BRATS 2017 challenge \cite{menze2014multimodal,bakas2017segmentation,bakas2017advancing}. Nevertheless, the ensemble with seven 3D architectures requires higher computation complexity and training time. 

\par
\hspace{0.5cm}In, \cite{chen2018mri}, Chen et al. proposed a hierarchical 3D CNN architecture to segment Glioma from multi-modal brain MRI volumes. The model uses two distinct scales of 3D image patches to examine multi-scale features. The 3D CNN architecture uses a series of hierarchical dense convolutions without pooling layers. The model uses a patch-wise analysis that reduces the class imbalance and the computational cost involved in training large 3D datasets. This method was validated on the BRATS 2017 \cite{bakas2017segmentation} dataset and reported better segmentation performance compared to similar patch-wise 3D approaches. 

\par
\hspace{0.5cm}The patch-based segmentation helps to reduce the hardware resources for training and generates a large number of data samples that favor adequate learning. However, the patch-wise analysis often fails to extract global features from the actual image volumes. This may limit the learning performance when the abnormality is region-specific. 

\item \textit{Multi-task learning models}

Multi-task learning (MTL) is a machine learning approach that can assess multiple related tasks with a single network \cite{vafaeikia2020brief}. MTL is advantageous in medical images because many tasks such as classification and segmentation may be required concurrently throughout the diagnostic process \cite{wu2021elnet}. Graphical representation of a 3D CNN based MTL is shown in Figure \ref{fig:MT}. This section discusses a few such approaches that use 3D CNN for multi-task learning.

\begin{figure} [!tbh]
\centering
\includegraphics[width=0.75\textwidth]{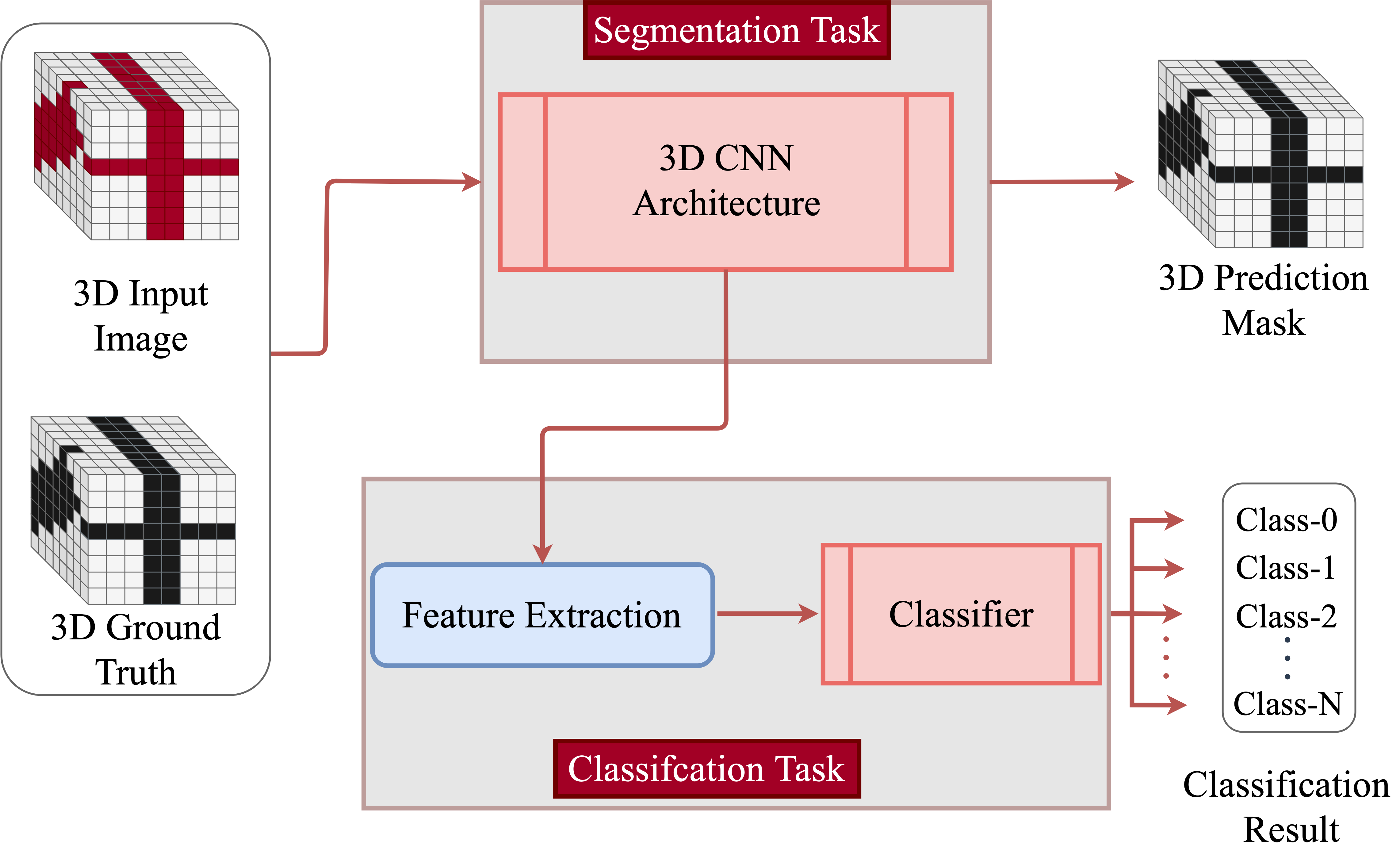}
\caption{A high-level block diagram of 3D CNN based Multi-Task learning model.}
\label{fig:MT}   
\end{figure}

\par
\hspace{0.5cm}Zhou et al. \cite{zhou2021multi} proposed multi-task learning of classification and segmentation using a 3D CNN model for classifying tumors in breast ultrasound images. The authors used a modified version of 3D V-Net \cite{milletari2016v} as the backbone of this model. To conduct a multi-scale analysis, feature maps from three higher-level encoder and decoder layers of the segmentation network are fused together to perform the classification task. Sharing higher-level feature maps and a multi-task loss function encourages the network to learn common generic features for classification and segmentation.

\par
\hspace{0.5cm}Another multi-tasking approach has been reported by Gordaliza et al. \cite{gordaliza2019multi} to infer tuberculosis from CT images. The model uses higher-order encoder features and processes two individual feed-forward neural networks to understand the tuberculosis condition. The approach also used several optimizations such as self normalization, the use of scaled exponential linear unit (SELU) activation function, and uncertainty weighted multi-task loss to improve the performance both for detection and counting the number of nodules.

\par
\hspace{0.5cm}Ge et al. \cite{ge2019k} proposed a multi-task learning approach for segmenting and quantifying left ventricle (LV) from paired apical views (A4C and A2C) of echo sequences. The method offers an overall cardiac analysis using a multi-task network: K-Net, an end-to-end network that can simultaneously segment LV and quantify its extent over the 3D plane. The methodology uses 2D convolution in different stages to segment and quantifies the 3D structure of LV using complex echo data. The reported results over a sufficiently large echo dataset also prove the proposed model's superiority in the heterogeneous learning approach.

\end{enumerate}

\subsubsection{Semi-supervised Learning}

\begin{figure} [!tbh]
\centering
\includegraphics[width=0.75\textwidth]{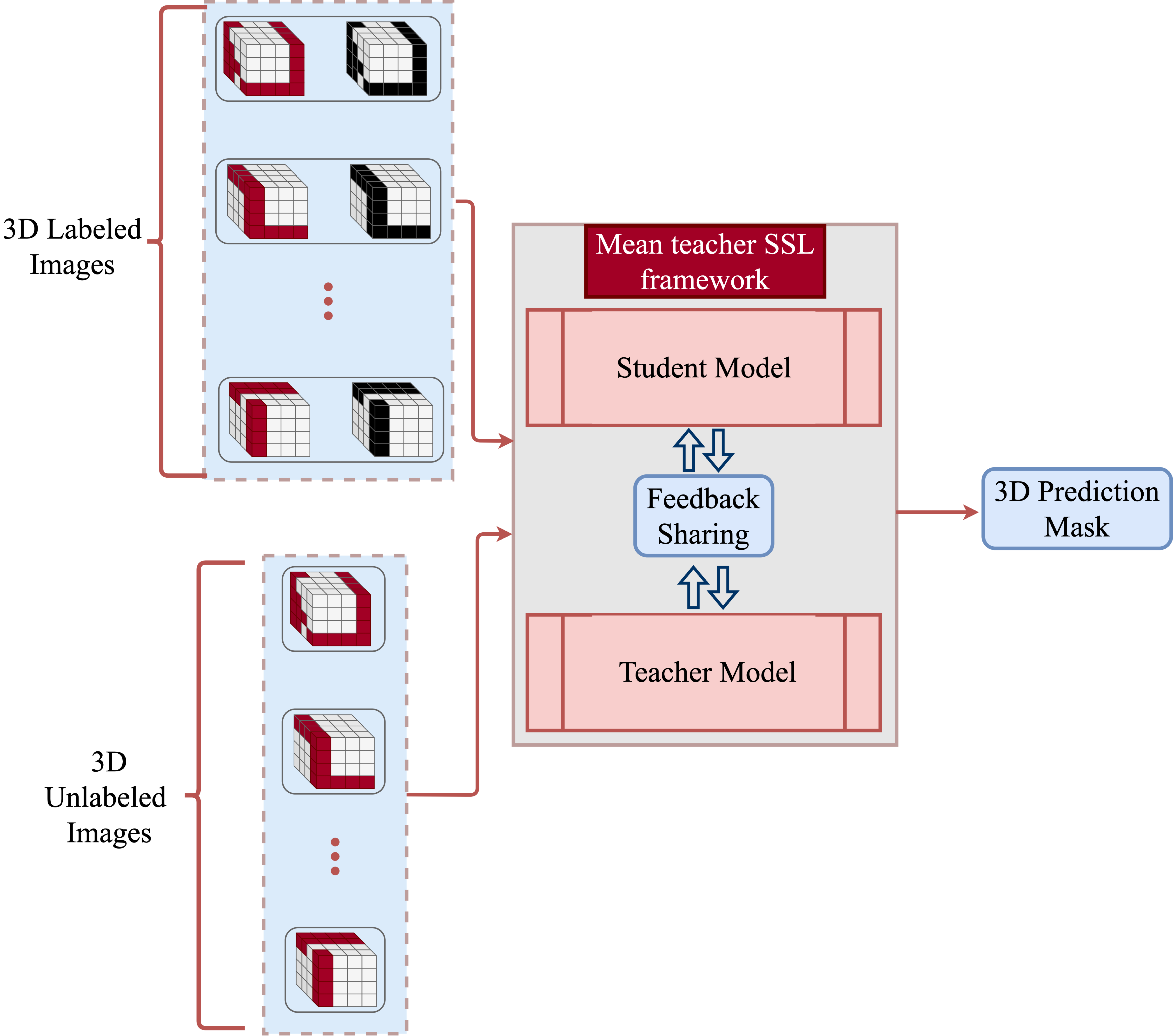}
\caption{A high-level representation of a 3D CNN based Mean teacher Semi-supervised learning model.}
\label{fig:SSL}   
\end{figure}

The 3D CNN models discussed in Section \ref{1} mostly use fully supervised deep learning algorithms that require thousands of annotated 3D volumes. However, accurate marking of ground-truth images is a labor-intensive and tedious process. Hence, the supervised learning algorithms are more expensive in terms of both time and cost. Consequently, research also commenced on alternatives to process sparsely annotated training data. Semi-supervised learning (SSL) methods \cite{zhu2009introduction,rasmus2015semi,snell2017prototypical} are one of those types that require a few labeled image samples with a large number of unlabeled samples, and such SSL based 3D medical image segmentation works are discussed in this section. A basic 3D SSL framework is shown in Figure \ref{fig:SSL}.

In \cite{cciccek20163d}, {\c{C}}i{\c{c}}ek et al. presented one of the first promising 3D CNN network for volumetric segmentation that learns from sparsely labeled 3D images. The proposed model extends the classical U-Net architecture \cite{ronneberger2015u}  by substituting all 2D operations with their 3D equivalents. The study in \cite{cciccek20163d} outlined two cases: a semi-automated model and a fully automated model. In both cases, the network learns from  sparse annotated data and this helps to reduce the human effort in ground truth labeling without considerable degradation in the segmentation performance. The experiments were conducted on the Xenopus kidney dataset \cite{nieuwkoop1994normal}, and both the semi-automated and fully automated models showed considerable performance improvements over state-of-the-art 2D CNN architectures.

\par
In \cite{Mondal2018Fewshot3M}, Mondal et al. proposed a 3D multi-modal medical image segmentation method based on generative adversarial networks (GANs) \cite{Goodfellow2014GenerativeAN}. The method uses a semi-supervised training with a mix of labeled and unlabeled images. The proposed architecture uses several design considerations to modify the standard adversarial learning approaches to generate 3D volumes of multiple modalities. The 3D GANs generate fake samples and are used along with labeled and unlabeled 3D volumes, and a discriminator defines separate loss functions for these labeled, unlabeled and synthetic (fake) training samples. However, generating useful synthetic samples may not represent the actual data distribution and thus becomes challenging working with 3D medical image data. \par

Zhou et al. \cite{zhou2019semi} proposed a straight-forward semi-supervised segmentation approach named deep multi-planar co-training (DMPCT), which uses parallel training to extract information from multiple planes. The multiplanar fusion generates reliable pseudo labeling to train deep segmentation networks. The DMPCT framework consists of a teacher network that uses fully labeled images for training. The trained model then creates the pseudo labels from the unlabelled training data with a multi-planar fusion module. Subsequently, the student model uses both labeled and pseudo labeled data for the final training process.\par
Yang et al. \cite{yang2020deep} proposed a similar semi-supervised segmentation technique to detect catheter from volumetric ultrasound images. The segmentation model uses a deep Q network (DQN) for localizing the target region. After the catheter localization, the method uses a twin-UNet model for the semantic segmentation of the catheter volume around the localized region by a patch-based strategy. This model uses a typical teacher network followed by a student network to train both labeled and unlabeled 3D patches based on a set of hybrid constraints. 

\par
In \cite{li2020shape}, Li et al. presented a shape-aware 3D segmentation for medical images to use extensive unlabeled data so as to enforce a geometric shape analysis on the segmentation output. The model uses a deep CNN architecture that predicts semantic segmentation and signed distance map (SDM) of object surfaces. The network uses an adversarial loss between the predicted SDMs of labeled and unlabeled data during training to leverage shape-aware features. The integration of adversarial loss, which uses a generative discrimination function, helps supervise learning with unlabelled data and extract generalized features. 

\par
In \cite{wang2020focalmix}, Wang et al. proposed a tailored modern semi-supervised learning (SSL) method named as FocalMix for the detection of lesions from 3D medical images. The model is built on MixMatch  \cite{berthelot2019mixmatch} SSL framework, which uses a  prediction for unlabeled images and MixUp augmentation. The proposed 3D CNN model uses a Soft-target Focal Loss along with an anchor level target prediction to improve lesion detection. The study also uses two adaptive augmentation methods: image-level MixUp and object-level MixUp, to generate the final training data. 

\par
Zhang et al. \cite{zhang2021dual} proposed a 3D medical image segmentation model using semi-supervised 3D CNN. This dual-task mutual learning model uses two side-by-side frameworks. One network works on the region-based shape constraint, while the learning in the other network focuses on boundary-based surface mismatch. The main contribution of this model is the use of a signed distance map (SDM) and the conventional ground truth maps to get a better intuition of region features and shape features together. Both networks use an identical 3D V-Net architecture to learn the voxel characteristics with a shared loss function. During the training with labeled image volumes, the loss function concentrates more towards the difference in segmentation map and actual ground truth. On the other hand, while training with unlabeled images, the model uses a consistency loss function based on distance maps to ensure prediction consistency while working with similar images. Hence, the dual network tries to reduce both losses and aids a better segmentation accuracy.

\par
Another semi-supervised method is presented by Li et al. \cite{li2021hierarchical} which uses a 3D CNN-based mean teacher framework with hierarchical consistency regularization for 3D left atrium MR images. The model felicitates the prediction consistency between the teacher and student network at multiple scales. During training, the student network uses multi-scale deep supervision while hierarchically regularizing the network's prediction consistency. Hence, the model learns from both the labeled and unlabeled volumes by minimizing the supervised loss and consistency loss concurrently.

\par
In semi-supervised approaches, the training demands a subset of data with accurately marked ground truths. Usually, multiple networks are used in semi-supervised models to process the labeled and unlabeled data using a shared feedback mechanism. The learning is often synchronized by evaluating the segmentation losses and consistency in predicting unlabeled data. The performance of a semi-supervised model highly depends on this feedback and the accuracy of the annotated images. Since SSL requires precise annotation (for a subset of data), it is challenging while dealing with large 3D image datasets. Hence, several algorithms that can learn from weakly labeled data have also been published in the medical image segmentation domain. Those methods are discussed in the below section.

\subsubsection{Weakly-supervised Learning}
\label{weakly}

\begin{figure} [!tbh]
\centering
\includegraphics[width=0.75\textwidth]{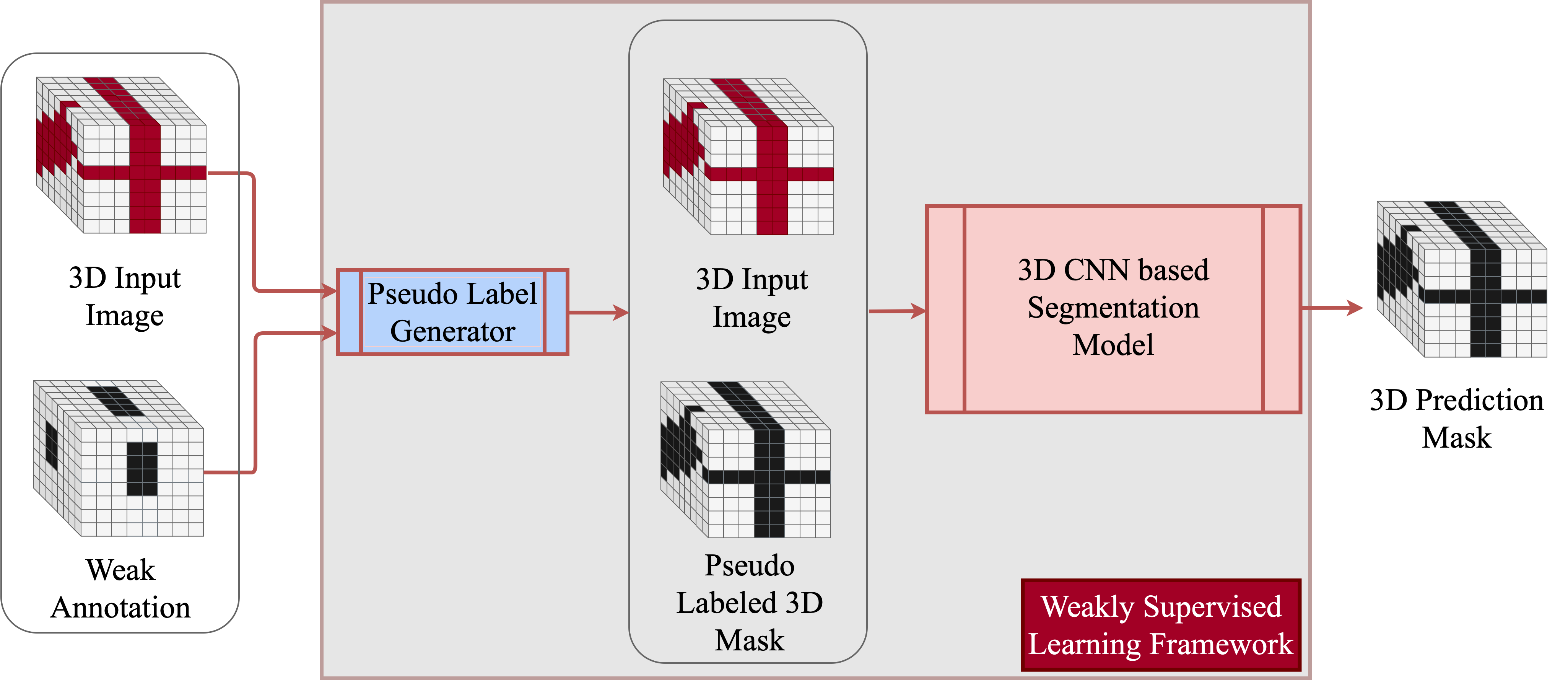}
\caption{Block diagram of a Weakly-supervised 3D CNN model.}
\label{fig:WSL}   
\end{figure}

Weakly supervised is a learning paradigm that uses noisy or low-quality annotations in the learning process. In weakly-supervised learning, the data labeling need not be as highly accurate as in the case of fully supervised learning. In medical image segmentation, weakly-supervised learning is highly significant as the abstract level annotation is relatively easier and may be accomplished by non-experts with minimal support from radiologists. The labels for weakly-supervised learning can also be made from batch clustering or from noisy predictions from comparable pre-trained models. Hence, the data preparation becomes relatively cheaper and practical, but at the cost of more noise or false labeling in the training data. A block level representation of a weakly-supervised 3D CNN is shown in Figure \ref{fig:WSL}. Several 3D CNN models with weak supervision have been reported recently, and those works are discussed here.

\par
Yang et al. \cite{yang2020weakly2} proposed a weakly-supervised method for segmenting catheters from 3D frustum ultrasound images. The methodology use data annotated with 3D bounding boxes over the catheter regions. A pseudo label generator module is introduced here to reduce the impact of inaccurate ground truth marking. This model uses a sequential network with a localization module to detect frustum volumes, followed by the segmentation stage to extract catheter voxels. The localization network used 3D ResNet-10 encoder architecture, and the feature maps are then converted to the final segmentation map using a decoder with multiple transpose convolutions. This approach makes the frustum segmentation faster and cheaper without any significant degradation in the segmentation performance.

\par
Wu et al. \cite{wu2019weakly} proposed a weakly-supervised brain lesion segmentation using attentional representation learning from 3D image volumes with image-level annotation. The approach used an attention mechanism that is dimensionally independent on the class activation map (CAM) and can estimate the lesion labels with minimum demands of the trainable parameters and learn the representation model from the dimensional independent CAM to extract the foreground voxels.

\par
Another similar weakly-supervised 3D CNN-based segmentation is proposed by Zhu et al. \cite{zhu2021weakly} that requires only image-level class labels. The proposed model: CIVA-Net, uses weakly annotated labels for volumetric image segmentation on 3D cryo-ET datasets. The input to the network is image-level class labels, and a pre-processing seed generation stage is used initially for generating pseudo labels for each voxel. Using the \emph{cross-image consensus} stage, the similar voxel groups are merged using co-occurrence learning to generate the pseudo localization map. An inter-voxel affinity learning is also proposed to analyze the inter-pixel relationship from the pseudo localization map to forge the affinity voxel pairs. The final segmentation stage uses VoxResNet \cite{chen2016voxresnet} to predict the segmentation map using the pseudo localization map and the affinity pairs generated from previous steps.

\par
Weakly-supervised learning is highly recommended for 3D medical images where the voxel shares a similar pattern in the region of interest. However, the pseudo labeling of the region of the region of interest just from the image class label is highly susceptible to errors and can reduce the segmentation performance significantly.

\subsubsection{Cost-effective approximations of 3D CNN models}

\begin{figure} [!tbh]
\centering
\includegraphics[width=0.75\textwidth]{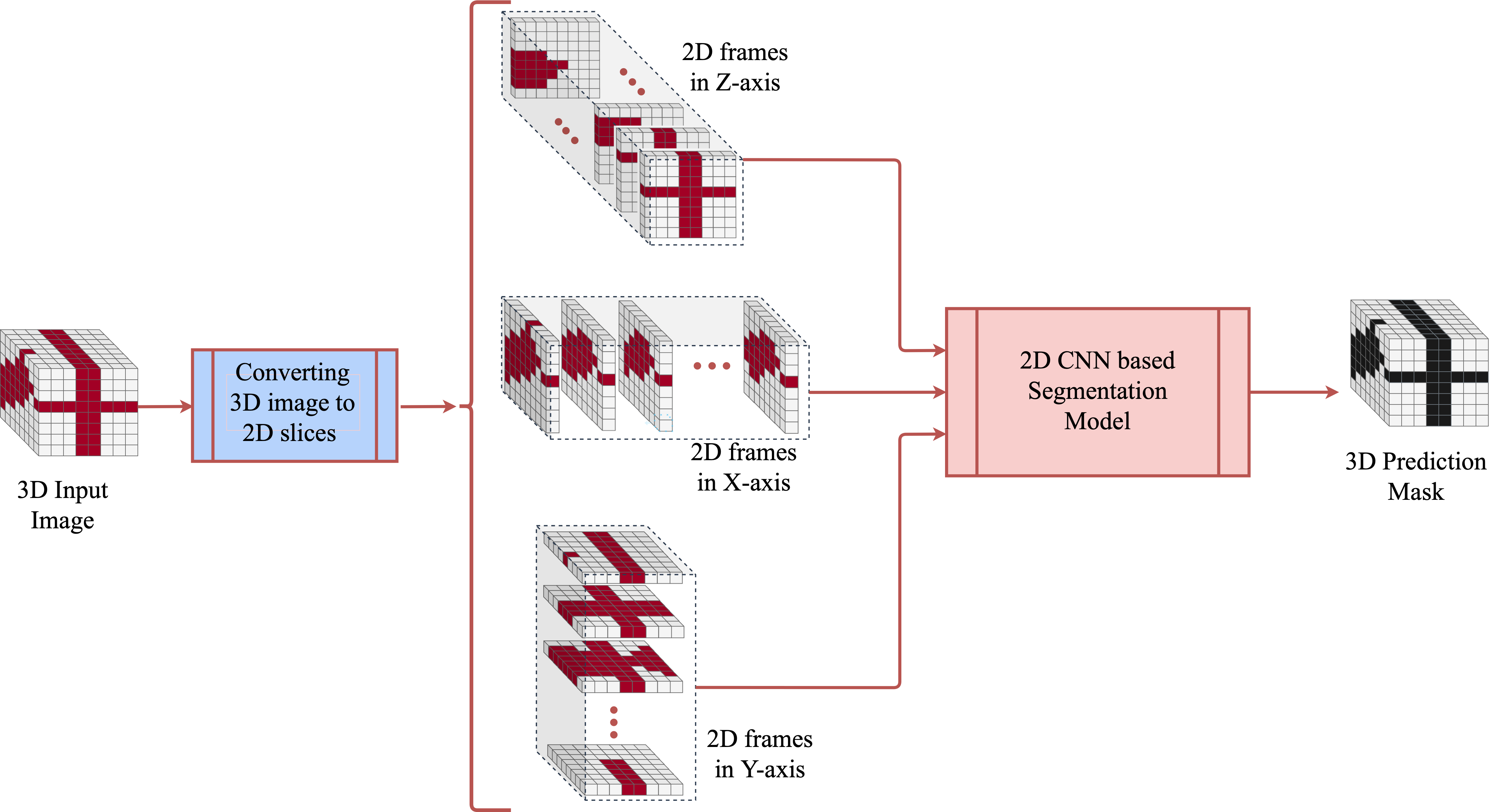}
\caption{A cost-effective approximation of 3D CNN model.}
\label{fig:Approx}   
\end{figure}

Since the straight-forward 3D CNN models are highly susceptible to large computation costs and require large datasets, several alternative techniques have been proposed that can use simulated 3D architectures with cost-effective approximations. For instance, 2D frames from a 3D image can be processed in three orthogonal directions and then merged to create the final 3D prediction mask, and such a model is shown in Figure \ref{fig:Approx}. Numerous prominent approximations for segmenting 3D medical images have been reported and discussed in this subsection.

\par
Extraction of inter-slice features from a 3D volume is possible by analyzing three orthogonal planes (sagittal, coronal, and transverse planes) and by classifying the voxel at the intersection of three planes. This method was successfully applied by Ciompi et al. \cite{Ciompi2017TowardsAP} for classifying lung nodules. This model learn from 2D patches in three perpendicular planes centered at a given voxel, at three different scales. Fully connected layers combine the streams and perform the voxel classification. However, CNN-based models with fully-connected layers are computationally less efficient for segmentation than FCNN architectures. A similar 3D segmentation approach by combining information from three orthogonal planes was discussed by Kitrungrotsakul et al. \cite{Kitrungrotsakul2019VesselNetAD}. The proposed method (VesselNet) uses 2D DenseNet \cite{Huang2017DenselyCC} network for classifying voxels in three different 2D planes. The features from the individual networks are then fused for getting the probability to predict the segmentation map. However, this approach is not an end-to-end segmentation model and hence misses most contextual and location-based information. 

\par
Mlynarski et al. \cite{Mlynarski20193DCN} proposed a similar CNN architecture that combines the benefits of the short-range 3D context and the long-range 2D context. This architecture uses modality-specific sub-networks to focus on missing MR sequences. During training, three individual 2D U-Net models were used to create features from axial, coronal, and sagittal slices. The final 3D network was then trained using these 2D features along with 3D patches from the input volumes. The study also suggests considering the outputs from multiple 3D models to minimize the constraints of specific choices of neural network models. However, the use of multiple 2D and 3D models demands more hardware resources, and patch-wise analysis fails to learn region-specific features. 

\par
Another 3D CNN model was introduced by Heinrich et al. \cite{Heinrich2019OBELISKNetFL}, which uses trainable 3D convolution kernels that learn both filter coefficients and spatial filter offsets in a continuous space, based on the principle of differentiable image interpolation. The proposed One Binary Extremely Large and Inflecting Sparse Kernel (OBELISK) works with fewer parameters and reduced memory consumption. The deformable convolutions in the OBELISK filter replace the continuously sampled spatial filter with a sparse sampling approach. This helps to extract information from wider spatial contexts and replace multiple small filter kernels at different scales. However, the computation complexity may be higher in OBELISK due to the unoptimized filter sampling. 

\par
Roth et al. \cite{roth2018spatial} proposed an automated segmentation system for 3D CT pancreas volumes based on a dual-stage cascaded method. It included a localization method followed by a segmentation. 2D holistically nested convolutional networks (HNN) \cite{xie2015holistically} on the three orthogonal directions, were used in the localization stage. The 2D HNN pixel probability maps are then merged to get a 3D bounding box of the foreground regions. In the second stage, the model focuses on semantic segmentation over the voxels in the bounding box. Two different HNN realizations are integrated in the segmentation stage that extract mid-level cues of deeply-learned boundary maps of the target organ. The authors also presented an advanced multi-class segmentation model \cite{roth2018application} for segmenting the liver, spleen, and pancreas from CT volumes. The cascaded network helps to provide boundary preserving segmentation and reduces false detection in the 3D volumes.

\par
Chen et al. \cite{Chen2018S3DUNetS3} presented a separable 3D U-Net architecture that targets to extract spatial information from the image volumes with limited memory and computation cost. The model uses a U-Net architecture for the end-to-end training with separable 3D (S3D) convolution blocks. The S3D block is an arrangement of parallel 2D convolutions and exploits the advantages of the residual inception architecture. The model is evaluated on BRATS 2018 \cite{bakas2018identifying} data, and the results justify the improvement in the segmentation performance compared to a standard 2D or 3D U-Net architecture. 

\par
Another 3D CNN architecture proposed by Rickmann et al. \cite{rickmann2020recalibrating} incorporates a compress-process-recalibrate (CPR) pipeline using 3D recalibration methods. The method uses a project \& excite (PE) module that compress the intermediate high dimensional 4D feature maps into three 2D projection vectors. The convolution layers in the \emph{processor} module learns from this 2D projections and the final recalibration stage generate the 4D tensors for the subsequent layers. This approach reduces the computation cost, without degrading the segmentation performance. The paper reported improved overall accuracy over different multi-class segmentation datasets \cite{roy2018recalibrating,jimenez2016cloud,jack2008alzheimer,kennedy2012candishare}. 

\par
The volumetric medical segmentation using the above-discussed approaches is highly useful in reducing the computation complexity and the need for large datasets. The results in the above discussed works prove its supremacy over other conventional 2D and 3D CNN approaches. However, the 3D approximations are highly dependent on the characteristics of the input data and the type of segmentation task. This may limit the generic use of models across different medical image modalities, and the designing of such a universal model remains an open research challenge.

\par
Some pre-trained 3D medical segmentation models have also been discussed in the literature. Chen et al. \cite{Chen2019Med3DTL} propose a heterogeneous 3D network called Med3D by co-training multi-domain 3D datasets to develop multiple pre-trained 3D medical image segmentation models. The authors use datasets from various medical challenges to create a 3D segmentation data set (3DSeg-8) with different modalities, target organs and pathologies. The pre-trained models were experimented on several 3D medical datasets \cite{Armato2011TheLI} using transfer learning to achieve performance gain and faster convergence. The model can work well on similar image modalities with less training time and can be considered as a suitable option for small 3D image datasets. Zhou et al. \cite{zhou2019models} proposed a similar set of pre-trained 3D CNN networks for classification and segmentation tasks in CT and MRI. The models are collectively known as Generic Autodidactic Models (Models Genesis), which uses learning by self-supervision. However, transfer learning may not be an appropriate solution in various scenarios due to the difference in image features across different imaging modalities and targeted abnormalities. 

\par
We summarize various medical image segmentation approaches using the 3D deep learning techniques in Table \ref{table_main}. The CNN architecture used, datasets, and remarks are also included in this table. A detailed discussion regarding the challenges in 3D medical image segmentation, possible solutions, and future directions are discussed in Section \ref{section4}.

\begin{adjustwidth}{1em}{1em}
\begin{center}

\centering
\begin{longtable}{p{3.0cm}p{4.5cm}p{7.5cm}}

\centering
\textbf{Reference}  & \textbf{Dataset} & \textbf{Remarks}\\
\endfirsthead
\multicolumn{3}{c}
{\tablename\ \thetable\ -- \textit{Continued from previous page}} \\
\hline
\textbf{Reference} & \textbf{Dataset} & \textbf{Remarks}\\
\hline
\endhead
\hline \multicolumn{3}{r}{\textit{Continued on next page}} \\
\endfoot
\hline
\endlastfoot

\multicolumn{2}{c}{\textbf{ \emph{Direct 3D CNN models}}} & \\

Milletari et al. \cite{milletari2016v} &  PROMISE2012 MRI dataset \cite{litjens2014evaluation} & 3D V-Net with residual blocks and strided convolution.\\
Bui et al. \cite{bui20173d}  &  Infant brain MRI (iSeg \cite{wang2019benchmark}) & D FCNN with dense blocks and strided convolution.\\
Kayalibay et al. \cite{kayalibay2017cnn} & Hand MRI \cite{gustus2012human}, BRATS 2013 \cite{BRATS2013}, and BRATS 2015 \cite{menze2014multimodal} & 3D U-Net with deep supervision. \\
Dou et al. \cite{dou20173d} & CT Liver (SLiver07 \cite{heimann2009comparison}) \& Heart MRI (HVSMR \cite{pace2015interactive}) & 3D U-Net with deep supervision and CRF \cite{zheng2015conditional} based post-processing.  \\
Li et al. \cite{li2017compactness} &  Brain MRI (ADNI \cite{wyman2013standardization}) & 3D Res-Net with dilated convolution. \\
Chen et al. \cite{chen2018voxresnet}  &  Multi-modal Brain MRI (MRBrainS \cite{mendrik2015mrbrains})& VoxResNet: a multi-modal dataset to learn from small datasets.  \\
Niyas et al. \cite{niyas_3DFCD}  & Brain MRI (Private dataset)& 3D Residual U-Net with shallow sliced stacking for generating 3D samples.  \\
Schlemper et al. \cite{Schlemper2018AttentionUL} &  Abdominal CT datasets \cite{p26,Roth2017Hierarchical3F} & Attention U-Net that can learn region specific features.  \\
Wang et al. \cite{Wang2019RPNetA3}  & Brain MRI (CANDI \cite{Kennedy2011CANDIShareAR}, and IBSR \cite{Rohlfing2004EvaluationOA}) & 3D U-Net with pyramid pooling \cite{He2015SpatialPP} for fusing feature maps from different decoding levels.   \\
Peng et al. \cite{Peng2020MultiScale3U} & Brain MRI (BRATS 2015 \cite{menze2014multimodal}) &  Multiple U-Net architectures with Xception \cite{Chollet2017XceptionDL} blocks. \\
Zhou et al. \cite{Zhou2020OnePassMN} & Brain MRI (BRATS 2015 \cite{menze2014multimodal}, BRATS 2017\cite{bakas2017segmentation,bakas2017advancing}, and BRATS 2018 \cite{bakas2018identifying}) & OM-Net integrated with a cross-task guided attention (CGA) module.\\

\multicolumn{2}{c}{\textbf{ \emph{3D Patch-wise Segmentation}}} & \\
 Yu et al.\cite{yu20163d}  & Heart MRI (HVSMR \cite{pace2015interactive}) & 3D Fractal network with deep supervision. \\
 Kamnitsas et al. \cite{kamnitsas2016deepmedic} & Brain MRI (BRATS 2015 \cite{menze2014multimodal} and BRATS 2016 \cite{bakas2015glistrboost}) &  DeepMedic: Multi-scale 3D CNN with residual connections with 3D fully connected CRF. \\
 Kamnitsas et al. \cite{kamnitsas2017ensembles} & Brain MRI (BRATS 2017 \cite{bakas2017segmentation,bakas2017advancing}) & An ensemble of two DeepMedic models, three 3D FCNN models, and two 3D U-Net models. \\
 Chen et al. \cite{chen2018mri}  & Brain MRI (BRATS 2017 \cite{bakas2017segmentation,bakas2017advancing})  &  Multi-modal 3D CNN using 3D patches at two different scales. \\

  \multicolumn{2}{c}{\textbf{ \emph{Multi-task learning models}}} & \\
  
 Zhou et al. \cite{zhou2021multi} & 3D Breast Ultrasound (Private dataset)  &  Multi-task network for both classification and segmentation.\\
 
 Gordaliza et al. \cite{gordaliza2019multi} & Chest CT (Private dataset)  &  Multi-tasking using self normalized neural networks.\\
  
 Ge et al. \cite{ge2019k} & Echo images (A4C and A2C) (Private dataset)  &  Segmentation and quantification of multi-view echo sequence using K-Net.\\
 
 \multicolumn{2}{c}{\textbf{ \emph{Semi-supervised Learning}}} & \\
 
{\c{C}}i{\c{c}}ek et al.\cite{cciccek20163d} & Xenopus kidney dataset \cite{nieuwkoop1994normal} & 3D U-Net with sparsely annotated volumes.\\
 Mondal et al. \cite{Mondal2018Fewshot3M} & Infant brain MRI (iSeg \cite{wang2019benchmark} \& Multi-modal brain MRI (MRBrainS \cite{mendrik2015mrbrains}) &  The 3D GAN \cite{Goodfellow2014GenerativeAN} that generates synthetic samples to train with labeled and unlabeled 3D volumes\\
 Zhou et al. \cite{zhou2019semi} &  Abdomen CT (Private dataset) &  Deep multi-planar co-training (DMPCT). \\
 Yang et al. \cite{yang2020deep} & 3D heart ultrasound (Private dataset) & A localization (using deep Q network (DQN)), followed by semantic segmentation (using a dual U-Net).\\
 Li et al. \cite{li2020shape} & 3D heart MRI \cite{xiong2021global} & 3D V-Net with signed distance map (SDM).\\
 Wang et al. \cite{wang2020focalmix} & Thoracic CT scans \cite{setio2017validation,national2011reduced} & FocalMix: A model built on MixMatch  \cite{berthelot2019mixmatch} SSL framework.\\
 Zhang et al. \cite{zhang2021dual} & Atrial Segmentation Challenge MR dataset \cite{xiong2021global} & SSL framework that uses labeled images and signed distance map for accurate segmentation.\\
 Li et al. \cite{li2021hierarchical} & Atrial Segmentation Challenge MR dataset \cite{xiong2021global} & Mean teacher framework with hierarchical consistency regularization and multi-scale deep supervision.\\
 
  \multicolumn{2}{c}{\textbf{ \emph{Weakly-supervised learning}}} & \\
 
 Yang et al. \cite{yang2020weakly2} &  ex-vivo dataset (Private Dataset) & Use coarsely marked 3d data. The framework contains ResNet-10 based classification followed by segmentation.\\
 Wu et al. \cite{wu2019weakly} &  Brain MRI (BRATS 2017 \cite{bakas2017segmentation,bakas2017advancing} \& ISLES 2015 \cite{maier2017isles}) & Weakly-supervised learning using image-level class labeling.\\
Zhu et al. \cite{zhu2021weakly} &  3D cryo-ET Dataset \cite{guo2018situ} & CIVA-Net: an end-to-end model that segment foreground voxels from image-level class labeling.\\
 
 \multicolumn{2}{c}{\textbf{ \emph{Cost-effective approximations of 3D CNN models}}} & \\
 Ciompi et al. \cite{Ciompi2017TowardsAP} & CT Lung (MILD \cite{pedersen2009danish} dataset) & ConvNet: Multi-stream voxel classification in three orthogonal planes at three different scales. \\
 Kitrungrotsakul et al. \cite{Kitrungrotsakul2019VesselNetAD}  & CT Liver (Private dataset) & VesselNet: Voxel classification in three orthogonal planes. \\
 Mlynarski et al. \cite{Mlynarski20193DCN} & Brain MRI (BRATS 2017 \cite{bakas2017segmentation,bakas2017advancing}) & Used features from three orthogonal planes with 2D CNNs, and 3D image volumes.\\
 Heinrich et al. \cite{Heinrich2019OBELISKNetFL}  & CT datasets \cite{jimenez2016cloud,roth2015deeporgan} & OBELISK-Net that requires fewer trainable parameters. \\
 Roth et al. \cite{roth2018spatial} & Abdomen CT dataset \cite{roth2015deeporgan} & A localization (using Holistically-nested convolutional networks (HNN) on three orthogonal planes) followed by a semantic segmentation stage. \\
 Chen et al. \cite{Chen2018S3DUNetS3} & Brain MRI (BRATS 2018 \cite{bakas2018identifying}) & Separable 3D (S3D) U-Net: A parallel 2D convolutions with a residual inception architecture.\\
 Rickmann et al.\cite{rickmann2020recalibrating}  & Multiple 3D medical datasets \cite{roy2018recalibrating,jimenez2016cloud,jack2008alzheimer,kennedy2012candishare} & 3D ConvNet with project \& excite (PE) modules. \\
\caption{Overview of articles using 3D deep learning techniques for medical image segmentation.}
\label{table_main}
\end{longtable}
\end{center}

\end{adjustwidth}

\section{Discussion \& Conclusion}
\label{section4}

\subsection{Summary}

In this study, we reviewed nearly 40 articles on 3D medical image segmentation using deep learning techniques. The articles were classified into different categories according to the different 3D deep learning approaches used to solve the segmentation problem. We believe that deep learning significantly contributes to medical image segmentation with 3D medical images. Although 3D deep learning methods are considerably less in use than 2D approaches, the exponential increase in the publications related to 3D deep learning techniques in the past couple of years in medical image analysis is remarkable. Nevertheless, most of the 3D CNN-based segmentation approaches rely on traditional CNN architectures by updating the convolution kernel and pooling from 2D to 3D. This bypass is advantageous for easy implementation; however, it introduces several optimization issues and restricts the analysis of 3D data to its full potential. 

\par
Numerous studies have focused on solving the inherent limitations of 3D deep learning, such as the computation cost and huge memory requirements while processing large number of 3D samples. The need for large volumetric datasets is another major challenge in 3D deep learning-based medical image segmentation. The key aspects of such successful alternative approaches in 3D deep learning-based segmentation were also reviewed in this study. The following subsections provide a detailed overview of the unique challenges faced in 3D medical image segmentation and the future trends.

\subsection{Challenges in 3D Medical Image Segmentation}

Deep learning methods are associated with various challenges in medical image analysis, and the use of 3D CNN for volumetric medical image segmentation further increases the complexity. The primary challenge faced by the researchers in implementing effective 3D CNN models is the need for large training data. The use of picture archiving and communication system (PACS)-based tools significantly improved the medical image management and storage standardization. However, huge amounts of archived medical data often become worthless due to the lack of proper annotation and the inconsistency in data attributed to the heterogeneity in imaging systems. 

\par
Most of the traditional 2D and 3D CNN models require task-specific labeled data, generally obtained from domain experts. Working with 3D medical data involves a significant time investment, particularly during slice by slice annotation. Similar to other supervised machine learning approaches, subjectivity in labeling is another challenge faced during 3D segmentation. Multiple domain experts might be required to decode the information corresponding to large 3D datasets. This may also introduce labeling noise in the training data and reduce the efficiency of volumetric segmentation. 

\par
In medical image analysis, generalized feature representation is often limited by the intra-class variance and inter-class similarity in large 3D datasets. Variations in image acquisition systems, contrast variations, and noise are the major reasons for overlapping features among different classes that critically influence the decision-making in unseen test data. Another limiting factor in 3D medical image segmentation is class imbalance. The class imbalance may result in over-segmentation of classes with a high voxel share. Generally, 3D medical image segmentation is required for detecting abnormal or lesion regions from a scanned volumetric image. Hence, the ratio of voxels belonging to the abnormal and normal classes is extremely low and seriously affects the training process. The class imbalance problem may be more critical in multi-class segmentation cases. Training a 3D CNN model on a slarge volumetric dataset is also challenging as it demands large memory and high computation capability.

\subsection{Future Directions}

Most of the above-discussed challenges are inherent in the medical data and can not be refrained from the source. Moreover, practically it is impossible to work with clean data. Hence, many studies are being conducted to design deep learning models to circumvent such issues. Since 3D medical image segmentation is primarily affected by the unavailability of labeled datasets, several recent studies have proposed techniques that can learn from partially or coarsely annotated data. Semi-supervised and weakly-supervised learning are some popular among those methods. Furthermore, such partially-supervised models can reduce the impact of label error and subjectivity to a certain extent. However, such approaches are still in the early stage. Hence, developing sophisticated semi-supervised 3D CNN models for volumetric medical image segmentation is still an open challenge for the research community.

\par
Another possibility for improving the 3D segmentation performance is by reducing the huge class imbalance. Several attempts \cite{salehi2017tversky,xia2019exploring,hashemi2018asymmetric} have been made to reduce class imbalance by using asymmetric loss functions to provide custom weights for different voxel classes. Although several alternative approaches, such as selective sampling and adaptive augmentation, are available to reduce the class imbalance problem, such strategies fail for highly imbalanced datasets.Hence, there is high scope for methods that address the class imbalance in 3D medical data. 3D CNN can learn multi-dimensional features and contribute significantly to multitask learning due to its comprehensive learning pattern. GAN-based data synthesis is another solution to resolve several of the current 3D (or 2D) medical image segmentation challenges. GAN-based analysis might be extremely relevant if it can generate real-like medical image volumes. Hence, such models for generating synthetic images will be an interesting research topic in the future. We believe that this analysis will help the research community to get better insights into current trends, future scopes, challenges and thus formulate new ideas bridging the research gaps in 3D medical image segmentation.

\bibliography{Ref}

\end{document}